\documentclass[a4paper]{jpconf}
\usepackage{graphicx}

\usepackage{amsmath}
\DeclareMathAlphabet{\mathpzc}{OT1}{pzc}{m}{it}

\begin{document}
\title{Event-By-Event Initial Conditions for Heavy Ion Collisions }

\author{S Rose and R J Fries}

\address{Cyclotron Institute, Texas A\&M University, College Station, TX, 77843-1411, US}

\ead{zenth @ physics.tamu.edu}

\begin{abstract}
The early time dynamics of heavy ion collisions can be described by classical fields in an approximation of Quantum ChromoDynamics (QCD) called Color Glass Condensate (CGC). Monte-Carlo sampling of the color charge for the incoming nuclei are used to calculate their classical gluon fields. Following the recent work by Chen et al. we calculate the energy momentum tensor of those fields at early times in the collision event-by-event. This can then be used for subsequent hydrodynamic evolution of the single events.
\end{abstract}

\section{Introduction}
Quark Gluon Plasma (QGP) is a state of matter created by high energy heavy ion collisions at the highest energies achieved at the Relativistic Heavy Ion Collider (RHIC) at Brookhaven National Laboratory and at the Large Hadron Collider (LHC) at CERN. The QGP state is believed to be close to thermal equilibrium soon after the collision, and is therefore studied by using viscous hydrodynamics to evolve the system until the point of kinetic freeze-out. Hydrodynamics requires an initial condition (for $\tau=\tau_{th}$), which can be motivated by describing the initial collision in terms of interactions between the colliding nuclei's gluon fields, which at high energies may be treated classically.

For ultrarelativistic collisions one particular approach for calculating these initial conditions is Color Glass Condensate (CGC). Ultrarelativistic nuclei are approximated to follow trajectories on the light cone. In the CGC approach, the gluon fields are treated classically, created by sources whose fluctuations are frozen in time. The framework to include corrections to the classical picture has been explored, but will not be addressed here (see, for example, \cite{Dusling:2011rz}). In recent years, attempts have shifted from describing smooth, initial states representing average initial conditions to models which allow for fluctuations for a single event. A particularly successful implementation is the IP-Glasma event generator \cite{Schenke:2013df}.

\section{Background}

Calculation of the gluon distribution at small values of Bjorken $x$, for large nuclei has been argued by McLerran and Venugopalan (MV) to be possible in systems with very high charge densities using weak coupling \cite{McLerran:1994}. In such a case, the density of partons per unit area, $\mu$, provides a scale much larger than the intrinsic QCD scale, where the strong coupling constant becomes small. This effect should be more pronounced in very large nuclei ($\mu \sim A^{^1/_3}$).

We follow analytical results presented by Guyangyao Chen et al.\,\cite{Chen:2013ksa}\cite{Chen:2015wia}. Here, we begin with presumed charge densities, $\rho_1$ and $\rho_2$, in the two nuclei giving the current
\begin{equation}
J^+_1\left(x\right) = \delta\left(x^-\right)\rho_1\left(\vec{x}_\perp\right), \qquad J^-_1\left(x\right) = 0,
\end{equation}
for the particle traveling in the $x^+$ direction. The current for $\rho_2$ is set up analogously and there is no current in the transverse direction. To determine the gluon field in the forward light cone, we solve the Yang-Mills equations and continuity equation. This work uses a known recursive solution for the coefficients $A^\mu_{(n)}$ of an expansion of the gluon field $A^\mu$ in the forward light cone,
\begin{equation}
A^\mu\left(\tau, \vec{x}_\perp\right) = \sum^\infty_{n=0}\tau^nA^\mu_{(n)}\left(\vec{x}_\perp\right).
\end{equation}

\section{CGC Event Generator}
While we simulate the collision on a grid, our work is analytic in character as much as possible. Whereas the grid parameters corresponding to the size of the system and the point spacing generally have effects on a numerical solution, we ensure that the physical infrared and ultraviolet scales in the problem are safe from these grid scales. The semi-analytic character of our event generator  sets this work apart from the existing IP-Glasma event generator by Schenke et al. which is a numerical solver \cite{Schenke:2013df} \cite{Gale:2012fu}. However, we expect the results of both generators to be consistent, if the same average charge densities are used, up to times where our power series is convergent, roughly $\tau \sim \frac{1}{Q_s}$.

First, we must sample the charges in the nuclei. To describe the fluctuation of charges, we model the charge density as having Gaussian fluctuations with a strength $\mu$ and strong coupling $g$, or
\begin{equation}
P\left(\rho\right) = \frac{1}{\sqrt{2\pi g\mu}}e^{-\rho^2/2g\mu },
\end{equation}
where we assume the transverse density $\mu$ to be a Woods-Saxon integrated over the longitudinal coordinate, an assumption that can be generalized readily later to include, for example, the IP-Sat model. Figure \ref{fig:ChargeSamp} shows a typical result of charge sampling.

\begin{figure}[t]
\caption{A sample of $\rho$ for a single color and nucleus of radius $R=3$ fm. The resolution scale is $\lambda=0.2$ fm.}\label{fig:ChargeSamp}
\centering
\includegraphics[width=0.5\textwidth]{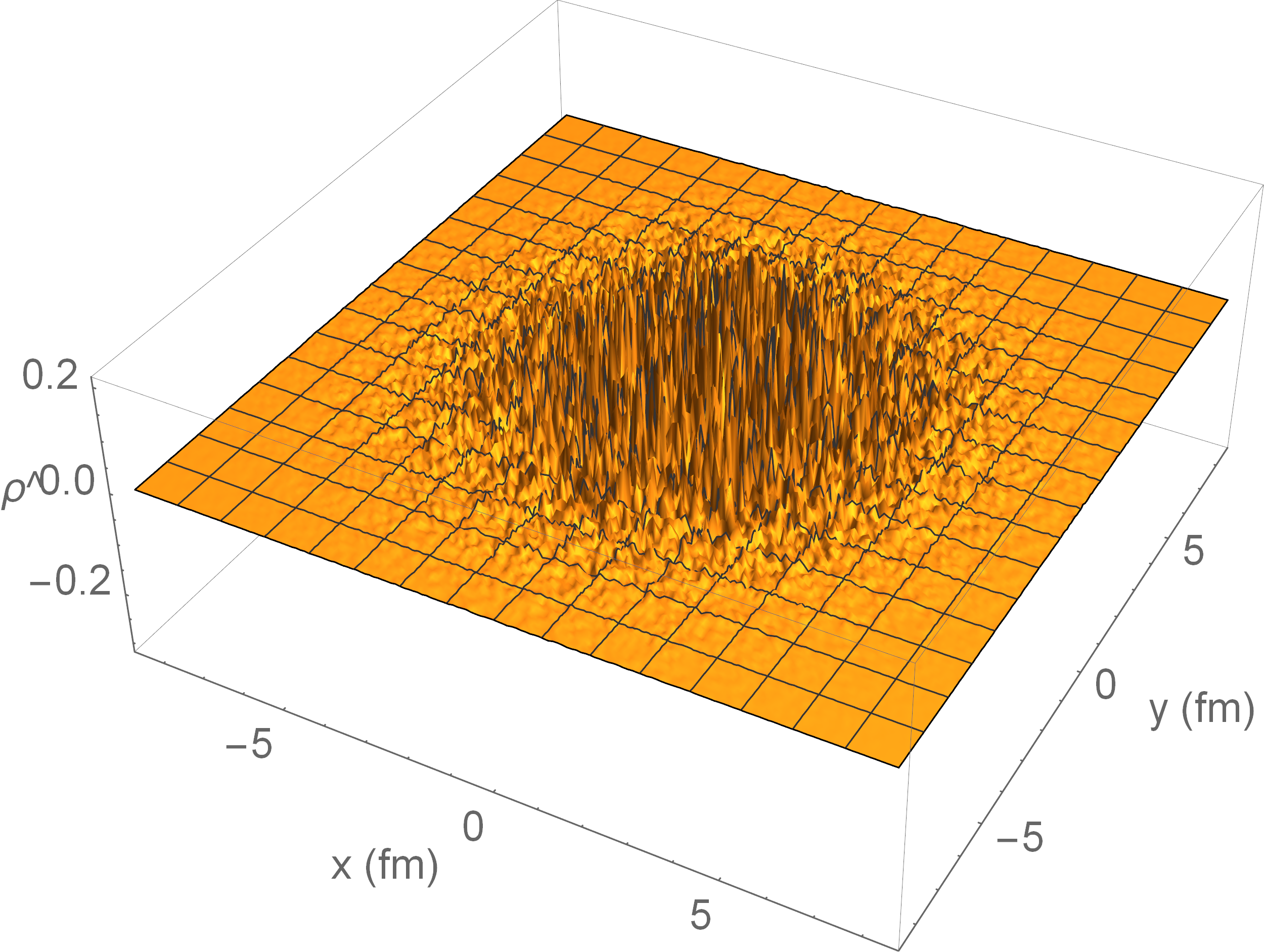}
\end{figure}

With the sampled $\rho$, we begin to solve for the gluon field. We initially solve in a covariant gauge where the gluon field is given by,
\begin{equation}
A^\mu_{cov}=\delta^{\mu\pm}\alpha.
\end{equation}
The benefit is that in this gauge, the Yang-Mills equations reduce to a Poisson equation,
\begin{equation}
\Delta\alpha\left(x^\mp,\vec{x}_\perp\right)=-\rho_{cov}\left(x^\mp,\vec{x}_\perp\right).
\end{equation}
A solution can be found with a regularized Green's function $G\left(x_\perp\right) =\frac{1}{2\pi}K_0\left(mx_\perp\right)$. Here, the effective gluon mass, $m$ acts as an infrared cutoff for the system. We then convolute this with a Gaussian coarse-graining function to implement a finite resolution, $\lambda$,
\begin{equation}
G_{C}\left(\vec{x}_\perp\right)=\int\frac{1}{2\pi}K_0\left(m\left(\vec{x}_\perp-\vec{z}\right)\right)e^{^{\left(\vec{x}_\perp-\vec{z}\right)^2}/_{\lambda^2}}dz.
\end{equation}
This approach lets us control the infrared and ultraviolet scale of the problem independently from the grid we simulate on. We find the field, $\alpha$, to be
 \begin{equation}
\alpha\left(\vec{x}_\perp\right) = \sum_{S} G_{C}\left(\vec{x}_\perp-\vec{z}_S\right)\rho\left(\vec{z}_S\right),
\end{equation}
where $S$ indicates the source locations, $\vec{z}_S$.

While the solution for $\alpha$ is straightforward, it is not a simple gauge for calculating observables. For this goal, we transform our solution into the light cone gauge via the Wilson Line,
 \begin{equation}
U\left(x_\perp\right)=\mathpzc{P}\exp{\left[-ig\int^{x^\pm}_{-\infty}\alpha\left( z^\pm,x_\perp\right)dz^\pm\right]}.
\end{equation}
The integration over $z^\pm$ indicates a requirement to sample the charge distribution multiple times along the light cone for even a single nucleus. We find then our transformed gauge field is,
 \begin{equation}
A^i\left(\vec{x}_\perp\right)=-\frac{1}{ig}U\left(\alpha,\vec{x}_\perp\right)\partial^iU^\dagger\left(\alpha,\vec{x}_\perp\right).
\end{equation}
Here, we must calculate $\partial^iU$ numerically, which we do using a finite difference scheme accurate to fourth order.

After the fields of the two nuclei are established we can now move on to calculate observables after the collision. We will start with the initial color-electric and color-magnetic fields. At $\tau=0$, we see that there are only longitudinal fields \cite{Chen:2015wia}
\begin{equation}
E_0=F^{+-}_{(0)}=ig\left[A^i_1, A^i_2\right], \qquad B_0=F^{21}_{(0)}=ig\epsilon^{ij}\left[A^i_1, A^i_2\right].
\end{equation}
Generally, the longitudinal components have non-zero terms for even orders, and transverse components have non-zero terms on odd powers of $\tau$.

Now we move on to the stress-energy tensor, which we will use for our initial condition. We can solve order-by-order in $\tau$. At $\tau=0$, only diagonal elements are non-zero,
\begin{equation}
T^{1 1}_{(0)}=T^{2 2}_{(0)}=\epsilon_0=-T^{3 3}_{(0)},
\end{equation}
Here, negative longitudinal pressure slows the participant remnants down, transferring energy into the glasma. The $1^{st}$ order contributions give initial flow by contributing to the Poynting vector,
\begin{equation}
T^{0 i}_{(1)}=\epsilon^{i j}\left(B_0 E^j_{(1)}-E_0B^j_{(1)}\right)=\frac{1}{2}\alpha^i\cosh \eta +\frac{1}{2}\beta^i\sinh\eta,
\end{equation}
\begin{equation}
T^{3 i}_{(1)}=E_0 E^i_{(1)}-B_0B^i_{(1)}=\frac{1}{2}\alpha^i\sinh \eta +\frac{1}{2}\beta^i\cosh \eta.
\end{equation}
We have introduced the radial flow,
 \begin{equation}
\alpha^i=-\nabla^i\epsilon_0,
\end{equation}
which drives the transverse expansion, and the directed flow,
 \begin{equation}
\beta^i=\epsilon^{i j}\left(\left[D^j,B_0\right]E_0 -\left[D^j,E_0\right]B_0\right),
\end{equation}
which originates from rapidity-odd transverse fields by an analogue to Gauss' Law. This analysis has been done to fourth order in work by G. Chen \cite{Chen:2015wia}. We have so far implemented up to second order in the event generator and plan to continue to higher orders (at least to fourth order). Figure \ref{fig:ChargeSamp} shows the initial energy density and flow field in the transverse plane for a typical event.

\begin{figure}[t]
\caption{Energy density and initial flow at $\tau = 0$ at $\eta \gg 1$ in a head on collisions for nuclei radius $R=3$ fm.}\label{fig:ChargeSamp}
\centering
\includegraphics[width=0.5\textwidth]{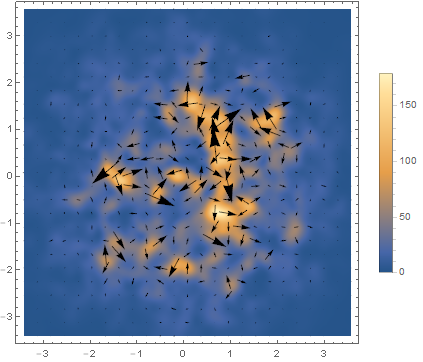}
\end{figure}

\section{Conclusion}

We have shown preliminary results from a CGC event generator currently under development. By utilizing analytic results, we will be enabled to investigate internal dynamics throughout the initial evolution of the system. A matching to viscous hydrodynamics, as in \cite{Gale:2012fu}, will take us further to kinetic freeze-out. In terms of observables, we have shown how directed flow may be calculated, which is anticipated to allow for angular momentum calculated from first principles to be put into the initial conditions for the first time.

\end{document}